\documentclass[%
reprint,
amsmath,amssymb,
aps
]{revtex4-2}

\usepackage[export]{adjustbox}
\usepackage{graphicx}
\usepackage{dcolumn}
\usepackage{bm}

\usepackage{color} 
\usepackage[colorlinks, linkcolor=blue, citecolor=blue, urlcolor=blue]{hyperref}

\newcommand{\cO}{\mathcal O}

\newcommand{\reef}[1]{(\ref{#1})}
\newcommand{\be}{\begin{equation}}
	\newcommand{\ee}{\end{equation}}
\newcommand{\bea}{\begin{eqnarray}}
	\newcommand{\eea}{\end{eqnarray}}
\newcommand{\ba}{\begin{equation}\begin{aligned}}
		\newcommand{\ea}{\end{aligned}\end{equation}}

\newcommand{\Df}{{\Delta_\phi}}

\begin{document}
	
	\preprint{APS/123-QED}
	
	\title{Polyakov blocks for the 1D CFT mixed correlator bootstrap}
	
	\author{Kausik Ghosh$^\Delta$}
	\email{kau.rock91@gmail.com}
	\author{Apratim Kaviraj$^r$}
	\email{apratim.kaviraj@desy.de}
	\author{Miguel F. Paulos$^\Delta$}
	\email{miguel.paulos@ens.fr}
	
	\affiliation{%
		{}$^\Delta$
		Laboratoire de Physique Th\'eorique,  de l'\'Ecole Normale Sup\'erieure, PSL University, CNRS, Sorbonne Universit\'es, \\UPMC Univ. Paris 06, 24 rue Lhomond, 75231 Paris Cedex 05, France\vspace{0.3cm}\\
		{}$^r$
		Deutsches Elektronen Synchroton DESY, Notkestrasse 85, 22603 Hamburg, Germany
	}%
	
	\date{\today}
	
	\begin{abstract}
		We introduce manifestly crossing-symmetric expansions for arbitrary systems of 1D CFT correlators. These expansions are given in terms of certain Polyakov blocks which we define and show how to compute efficiently. Equality of OPE and Polyakov block expansions leads to sets of sum rules that any mixed correlator system must satisfy. The sum rules are diagonalized by correlators in tensor product theories of generalized free fields. We show that it is possible to do a change of a basis that diagonalizes instead mixed correlator systems involving elementary and composite operators in a single field theory. As an application, we find the first non-trivial examples of optimal bounds, saturated by the mixed correlator system $\phi,\phi^2$ in the theory of a single generalized free field.

	\end{abstract}
	
	\maketitle
	\subsubsection{Introduction and setup}
	Conformal theories in one dimension are interesting both in theory and in practice. On the one hand, they have a wide range of applications: from conformal defects to boundary conditions in 2d CFT, passing through 2d QFTs in AdS space and long range spin models, not to mention that in a sense, any CFT is a 1d CFT \cite{Billo:2013jda, Gaiotto:2013nva, Cuomo:2021kfm, Cuomo:2021rkm,  Gimenez-Grau:2022czc,   Antunes:2021abs,Paulos:2016fap,Homrich:2019cbt,Knop:2022viy,Cordova:2022pbl}. On the other, being confined to a line and lacking spin, such theories have greatly simplified kinematics, while being far from elementary toy models, as there is no known non-trivial example which has been exactly solved. Thus these systems offer both a challenge as well as opportunity for significant progress in the conformal bootstrap program.
	
	At its heart this program is about understanding how crossing equations constrain sets of CFT data, both analytically and numerically. As it turns out, working with such equations in their original formulation in position space is far from optimal for the purpose of deriving such constraints. Instead, work in recent years suggests one should apply a certain transform from position space onto an auxiliary functional space \cite{Mazac:2018mdx, Mazac:2018ycv}. By construction, this mapping is done in such a way that crossing equations become translated into a completely equivalent set of sum rules which are now transparently solved by particular sparse sets of CFT data, something which is completely obscured in position space. Since all sets of CFT data, sparse or not, obey a measure of universality for large scaling dimensions \cite{Pappadopulo:2012jk,Qiao:2017xif}, these sum rules naturally give rise to a decoupling between low and high energy data, leading to rapidly convergent bounds and constraints \cite{Paulos:2019fkw}.
	
	Up to now constructions of such functional spaces have been limited to systems of correlation functions of operators lying in the same symmetry multiplet \cite{Ghosh:2021ruh}. This is a major restriction, as it is known that many CFTs of interest can only be effectively bootstrapped by considering systems of correlations functions involving distinct operator multiplets, the most famous example being the celebrated 3d Ising island \cite{El-Showk:2012cjh,El-Showk:2014dwa,Kos:2014bka}. In this paper we will resolve this shortcoming for 1d CFTs, by constructing new sets of sum rules valid for arbitrarily large systems of bootstrap equations. Our approach is to propose a generalization of the so-called Polyakov bootstrap to a multi-correlator setup \cite{Polyakov, Gopakumar:2016wkt, Mazac:2018mdx, Mazac:2018ycv, Gopakumar:2018xqi, Gopakumar:2021dvg,Ghosh:2021ruh}. Contrary to previous work, this allows us to bypass the labourious construction of functional kernels which implement the above mentioned transform, obtaining instead the relevant sum rules directly. The price to pay is that one cannot rigorously prove that these sum rules are really equivalent to the constraints of crossing. In the present work we will settle for checking that our sum rules pass several highly non-trivial consistency checks, giving us enough confidence to begin using them for both analytic and numeric explorations. In both cases we show they significantly outperform traditional approaches, leading to new qualitative and quantitative insights into the structure of crossing equations and the systematics of bootstrapping 1d CFTs.
	\vspace{0.3cm}
	
	{\em Setup ---} After these remarks, let us begin by recalling some basic facts and establishing notation. We are interested in 1d CFTs with primary operators labeled generically as $\Phi_i$. Conformal invariance dictates a correlation function $\langle \Phi_l(\infty)\Phi_k(1) \Phi_j(z) \Phi_i(0)\rangle$ can be expressed as a function $\mathcal{G}^{ijkl}(z)$ of a single cross-ratio $z$. 
	Using the OPE $\Phi_i\times\Phi_j$ (or simply $ij$), we obtain an expansion for $\mathcal{G}$:
	\be\label{Gope}
	\mathcal{G}^{ij,kl}(z)=\sum_{\cO}\lambda^{ij}_\cO \lambda^{kl}_{\cO}G^{ij,kl}_{\Delta_\cO}(z)\,.
	\ee
	The sum runs over all primary operators $\cO$ labelled by their scaling dimension $\Delta_\cO$ and spacetime parity $P_\cO=\pm$.
	Here $\lambda^{ij}_{\cO^\pm}=\langle \Phi_i(\infty)| \Phi_{j}(1) |\cO^\pm(0)\rangle=\pm \lambda^{ji}_{\cO^\pm}$ and $G^{ij,kl}_{\Delta_{\cO}}(z)$ is the 1d conformal block, see eq. (6) in supplementary materials \footnote{See  supplementary materials (which includes ref. \cite{DHoker:1999mqo, Dolan:2003hv, Zhou:2018sfz, Gopakumar:2016cpb,  Mack:2009mi, Mack:2009gy,  Mazac:2016qev}) for further details on Witten diagrams,   bootstrapping the system of decoupled GFF scalars and the $\phi,\phi^2$ system, and  numerical evidence of particle  production.}. 
	
	For a system of mixed correlators we find it useful to introduce the ``OPE orientation vector" $r^{ij}_\cO$, a new set of quantum numbers such that $\lambda^{ij}_\cO=|\lambda_\cO|r^{ij}_\cO$ and $\sum_{i,j}(r^{ij}_\cO)^2=1$, which describes how $\cO$ couples to different pairs of operators. We may then  rewrite \eqref{Gope} as
	\ba
	\mathcal G^{ij,kl}(z)=\sum_{\cO} \lambda^2_{\cO} G^{ij,kl}_{\cO}(z)\,,
	\label{eq:cftcorrelator}
	\ea
	with $G^{ij,kl}_{\cO}(z):=r^{ij}_\cO r^{kl}_\cO G^{ij,kl}_{\Delta_\cO}(z)$.
	
	An important property of a four-point correlator is crossing symmetry.
	The OPE decomposition given by~\reef{eq:cftcorrelator} is not manifestly crossing symmetric, as conformal blocks are associated to a particular OPE channel. To remedy this we introduce the {\em Polyakov block expansion}: \cite{Polyakov,Gopakumar:2016wkt,Mazac:2018ycv} 
	\be
	\mathcal G^{ij,kl}(z)=\sum_{\cO} \lambda^2_{\cO} \mathcal P^{ij,kl}_{\cO}(z).
	\ee
	By construction, the Polyakov blocks $\mathcal P_{\cO}^{ij,kl}$ are built to manifestly satisfy crossing. In particular, while in~\reef{eq:cftcorrelator} the only operators which give a non-zero contribution are those in the $s$-channel  OPE $\Phi_i\times\Phi_j$, the Polyakov block sum receives contributions also from the OPE channels $ik$ and $il$. The price to pay for this representation is that term by term the OPE contains contributions from unphysical states which must decouple in the full sum. Concretely, the {\em Polyakov bootstrap} is the statement:
	%
	\be
	\sum_{\cO}\lambda^2_{\cO}\left[G^{ij,kl}_\cO(z)-\mathcal P_{\cO}^{ij,kl}(z)\right]=0.
	\label{eq:polyboots}
	\ee
	This should be thought of as a reformulation of the constraints of crossing symmetry, which has to be satisfied by any system of correlators in any CFT. We will shortly show how these bootstrap equations may be turned into a more 
	useful discrete set of sum rules on the CFT data by using the OPE decomposition of the Polyakov blocks. 
	
	As functions, Polyakov blocks can be computed as sums of Witten diagrams. Starting off with the theory of $N$ decoupled free fields $\Psi_i$ in AdS$_2$, the associated boundary correlators correspond to those of the tensor product theory of $N$ generalized free fields (GFF) $\Phi_i$. Introduce now a new spin-0 bulk field $\chi_\Delta$ with mass $m^2=\Delta(\Delta-d)$ and dual operator $\cO_{\Delta}$ of positive parity~\footnote{Positive parity is reflected in the fact that $r^{\cO}_{ij}=r^{\cO}_{ji}$. In AdS$_2$ a negative parity operator on the boundary couples instead to a (massive) bulk spin-1 field, for which  $r^{\cO}_{ij}=-r^{\cO}_{ji}$.}, with couplings:
	\ba
	\mathcal L^{\mbox{\tiny int}}\propto \int_{\mbox{\tiny AdS}} \sum_{i,j}^N r_{\cO}^{ij}\, \chi_\Delta \Psi_i \Psi_j.
	\ea
	Then to leading order, the connected correlators in this theory are essentially proportional to the Polyakov block with the right quantum numbers, including the OPE orientation $r_\cO$. 
	We emphasize that this construction is simply a convenient recipe for computing the Polyakov blocks {\em as functions}: the constraints \reef{eq:polyboots} are meant to hold for all CFTs and not just those arising from weakly coupled fields in AdS$_2$.

	\vspace{0.3cm}
	
	\subsubsection{Sum rules}
	Since Polyakov blocks correspond to deformations of generalized free correlators,  their OPE content consists of double trace operators, whose schematic form and scaling dimensions are given as:
	\ba
	(ij)_n&\equiv   \Phi_i \Box^n\Phi_j\,,&\ \Delta_{(ij)_n}&=\Delta_i+\Delta_j+2n\, \\
	[ij]_n&\equiv \Phi_i \overset{\leftrightarrow}{\partial} \Box^n \Phi_j\,,&\  \Delta_{[ij]_n}&=\Delta_i+\Delta_j+2n+1\,.
	\ea
	At this point it is convenient to introduce some notational shorthand. We will denote a set of external fields $ij,kl$ by a letter $E$ (for {\em E}xternal). We will also denote by $I_c$ (for {\em I}nternal {\em c}hannel) a generic double trace operator appearing in the OPE channel $c$. A given $I_c$ is always with respect to some particular $E$. As an example, for $E=\{ij,kl\}$ we have
	\be
	I_{s}^+\in\mathcal I_{s}^{E,+}:=\{ (ij)_n\}_{n=0}^\infty \cup \{(kl)_n\}_{n=0}^\infty
	\ee
	Similarly we denote $I_{t}^{+}$ and $I_{u}^{+}$ for the double traces $(il)_n,(jk)_n$ and $(ik)_n,(jl)_n$ respectively. The negative parity channels are obtained from the above by swapping round and square brackets. We will also denote $I_s\in \mathcal I_{s}^{E,+}\cup \mathcal I_{s}^{E,-}$, and so on. Finally, we set
	\be
	r_\cO^{ij,kl;s}:=r_{\cO}^{ij}r_{\cO}^{kl}
	\ee
	and similarly $r_\cO^{ij,kl;t}=r_{\cO}^{il}r_{\cO}^{jk}$ and $r_\cO^{ij,kl;u}=r_{\cO}^{ik}r_{\cO}^{jl}$.
	%

	\vspace{0.1 cm}
	After these notational preliminaries we are ready to discuss the OPE for Polyakov blocks \cite{Note1}.
	As mentioned, they can be written as sums of Witten diagrams in AdS$_2$, which include both exchanges and contact diagrams. Exchange diagrams have conformal block decompositions as follows \cite{Gopakumar:2018xqi,Penedones:2010ue}:
	\ba \label{eq:Witten-deco}
	W^{E,c}_{\Delta,P}(z)=\delta_{c,s} G_{\Delta}^E(z)-\sum_{I_s} \alpha_{I_s}^{E,c}(\Delta,P) G^E_{\Delta_{I_s}}(z)
	\ea
	where parity $P$ of the exchanged operator (even,odd) corresponds to the bulk spin $\ell=0,1$ respectively. The coefficients appearing above satisfy the orthogonality properties 
	\ba\alpha_{I_s^\pm}^{E,c}(\Delta_{{I_c'}^\pm},P)=\delta_{c,s}\delta_{P,\pm} \delta_{I_s^\pm,{I_c'}^{\pm}}. \label{eq:duality}
	\ea
	%
	%
	As for contact diagrams, their block expansions take the form
	\ba
	W^{E,\text{con}}(z)&=\sum_{I_s}\alpha^{E,\text{con}}_{I_s}\, G_{\Delta_{I_s}}^E(z)\,.
	\ea
	The Polyakov blocks are then given as
	\ba\label{eq:Polyakov-block}
	\mathcal P^E_{\cO}(z)=\sum_{c} r^{E,c}_{\cO} W^{E,c}_{\Delta,P}(z)+ \mbox{contact diagrams}\,.
	\ea
	We will fix the contribution of contact diagrams momentarily. Using the OPE and commuting sums, the Polyakov bootstrap equations become a set of  conditons on  OPE data which can be written as follows:
	\begin{widetext} 
		{\bf Functional sum rules} --- For all $I_{s}\in\mathcal{I}_s^E$  there is a functional $\alpha^E_{I_s}$ with the sum rule 
		\begin{equation}
			\sum_{\cO}\lambda^2_{\cO}\alpha^E_{I_s}[\cO]:=\sum_{\Delta,P=\pm}\lambda^2_{\Delta,P}\left[ r^{ij,kl;s}_{\Delta,P}\, \alpha_{I_s}^{ij,kl;s}(\Delta,P)+ r^{ij,kl;t}_{\Delta,P} \,\alpha_{I_s}^{ij,kl;t}(\Delta,P)+ r^{ij,kl;u}_{\Delta,P} \, \alpha_{I_s}^{ij,kl;u}(\Delta,P)\right]=0\,. \label{eq:sumrules}
		\end{equation}
	\end{widetext}
	The coefficients appearing in these sum rules are the {\em functional actions}. They satisfy a set of duality properties following the orthogonality relations ~\reef{eq:duality}.
	%
	A slight issue arises when we have $\Delta_i+\Delta_j=\Delta_{k}+\Delta_l$, for instance in a correlator $\langle \phi_1 \phi_2 \phi_1 \phi_2\rangle$. In that case some coefficients in the Witten exchange diagram OPE expansion are degenerate \cite{Note1},  and we must make the replacements
	\ba
	\{\alpha_{(ij)_n},\alpha_{(kl)_n}\}&\to \{\alpha_{(ij)_n},\beta_{(ij)_n}\},
	\ea
	and similarly for $[ij]_n$.
	The duality properties  are then (writing $\alpha^E_{I_s}[I_c']:= \alpha_{I^{P'}_s}^{E,c}(\Delta_{{I'_c}^{P'}},P)$, etc) 
	\ba
	\alpha^E_{I_s}[I_c']&=\delta_{c,s}\delta_{I_s,{I_c'}}\,,\quad &\partial_{\Delta} \alpha^E_{I_s}[I'_c]=0\,,\\
	\partial_\Delta \beta^E_{I_s}[I'_c]&=\delta_{c,s}\delta_{I_s,I'_c}\,,&\beta^E_{I_s}[I'_c]=0\,.
	\ea
	The sum rules and duality relations have an implicit dependence on the contact diagrams as described below.
	Ignoring this part, there is one sum rule per label $I_s\in \mathcal I_s$ and per choice of $E$.
	
	Finally, let us discuss how to fix contact diagrams. Our guiding principle is that the sum rules  should bootstrap solutions to crossing with the same UV, or Regge behaviour, as ordinary CFT correlators \cite{Mazac:2018ycv,Ferrero:2019luz}. In AdS$_2$ such solutions can be contact diagrams arising from {\em relevant} deformations in the bulk theory i.e. four-point interactions with at most two bulk derivatives. So for each choice of external states we include all such independent contact diagrams in the Polyakov block~\eqref{eq:Polyakov-block}. We also require that we lose as many sum rules from~\reef{eq:sumrules} as there are contact diagrams. For example, for a certain choice of external states $E$ if there is only one independent contact diagram for all the possible permutations, then we would  redefine
	\ba
	\mathcal P_{\cO}^E(z)&\to \mathcal P_{\cO}^E(z)+\frac{\alpha^{\hat E}_{\hat I_s}[\cO]}{\alpha_{\hat I_s}^{\hat E,\text{con}}}\, W^{E,\text{con}}(z),\\
	\alpha_{I_s}^E[\cO]&\to \alpha_{I_s}^{E}[\cO]-\frac{\alpha^{E,\text{con}}_{I_s}}{\alpha^{\hat E,\text{con}}_{\hat I_s}} \alpha^{\hat E}_{\hat I_s}[\cO],
	\ea
	where $\hat E$ corresponds to some definite permutation of the indices in $E$. The above eliminates the functional $\alpha^{\hat E}_{\hat I_s}$, as well as the corresponding duality relations. This procedure is such that by construction, contact diagrams are now manifest solutions to the sum rules. For notational clarity, below we will leave the subtraction procedure implicit.

	\subsubsection{\label{sec:bounds}Basis change and optimal bounds}
	
	
	{\em The $\phi$, $\phi^2$ system --- } The duality relations \reef{eq:duality} imply that the functional sum rules trivialize on a set of correlators of decoupled GFF scalars \cite{Note1}. As a more interesting application, here we will examine how to bootstrap the $\phi,\phi^2$ system in a single GFF theory.  In our language, this is the following system of unitary CFT correlators of fields $\Phi_1$ and $\Phi_2$ with dimensions  $\Df\equiv \Delta_1=\frac 12\Delta_2$:
	
	\begin{equation}  \label{eq:gfffamily}
		\begin{split}
			\mathcal G^{11,11}(z)&=\mathcal G_{\Df}^{\mbox{\tiny gff}}(z),\, \mathcal G^{12,12}(z)=1+a^{12,12}_1\left(\mathcal G_{\Df}^{\mbox{\tiny gff}}(z)-1\right), \\ \mathcal G^{22,22}(z)&=\mathcal G^{\mbox{\tiny gff}}_{2\Df}(z)+\frac{a^{22,22}_2}2
			\frac {z^{2\Df}+(1-z)^{2\Df}+1}{z^{2\Df}(1-z)^{2\Df}}.
		\end{split}
	\end{equation}
	with $\mathcal G^{\mbox{\tiny gff}}$ the GFF correlator (see (25) in \cite{Note1}).
	There is a $Z_2$ symmetry under which $\Phi_1$ and $\Phi_2$ are odd $(-)$ and even  $(+)$ respectively. It is straightforward to expand the above in conformal blocks to extract the CFT data. 
	
	Let us discuss how the same data is reproduced using our sum rules. 
	We include in all OPE channels all possible single and double trace operators consistent with the symmetry.
	To allow for possible degeneracies we introduce the notation $a^{ij,kl}_{\cO_\pm}=\sum_{\Delta=\Delta_\pm}\lambda^{ij}_{\Delta_\pm}\lambda^{kl}_{\Delta_\pm}$. 
	We then must impose the non-degeneracy conditions:
	\ba
	a_{1}^{12,12}=a_{(11)_0}^{11,11}\,, \quad a_{2}^{22,22}=(a_{(11)_0}^{11,22})^2/a_{(11)_0)}^{11,11}\,. \label{eq:nondeg}
	\ea
	These tell us that there are unique operators with dimensions $\Delta_\phi$ or $2\Delta_\phi$ respectively. Note that these conditions cannot be derived, but rather must be imposed as inputs which fix the solution to be bootstrapped (i.e. there are solutions to crossing where they are not true \cite{Note1}).
	We can now determine all other OPE coefficients. In particular, the $\alpha_{(11)_n}^{11,22}$, $\alpha_{(12)_n}^{12,12}$ $\alpha_{[12]_n}^{12,12}$ sum rules determine the OPE coefficients  $a^{11,22}_{(11)_n}$, $a^{12,12}_{(12)_n}$ and $a^{12,12}_{[12]_n}$ respectively in terms of $a^{12,12}_{\Delta_\phi}$. Similarly, using  $\alpha_{(22)_n}^{11,22}$ the $a^{11,22}_{(22)_n}$ are also determined and come out zero as expected. At this point we use the nondegeneracy condition $a^{22,22}_{(11)_n}=(a^{11,22}_{(11)_n})^2/a^{11,11}_{(11)_n}$ and now the remaining OPE data can be solved for using the $\alpha^{11,11}_{(11)_n}$ and $\alpha^{22,22}_{(22)_n}$ sum rules.
	Our results match those extracted from the correlators above, giving a non-trivial check of our sum rules.\vspace{0.5cm}
	
	{\em Basis change --- }
	A dissatisfying feature of the last computation was that the basis of sum rules was not entirely diagonal with respect to the solution. 
	Indeed the equations in $22,22$ channel involve an infinite number of variables, and have to be solved after the $11,11$ and $11,22$ channels . Here we will construct a new functional basis that does not suffer from this problem. Consider the functionals $\alpha^{22,22}_{(22)_n}$. We will construct modified versions, $\widehat \alpha^{22,22}_{(22)_n}$ satisfying the following duality conditions:
	\ba
	\widehat \alpha^{22,22}_{(22)_n}[(22)_m]&=\delta_{n,m}\,,&     \quad \,\partial_{\Delta}\widehat \alpha^{22,22}_{(22)_n}[(22)_m]&=0\,,\\
	\widehat \alpha^{22,22}_{(22)_n}[(11)_m]&=0
	& \quad \partial_{\Delta}\widehat \alpha^{22,22}_{(22)_n}[(11)_m]&=0
	\,,&\quad \\
	\partial_{r}\widehat \alpha^{22,22}_{(22)_n}[(22)_m]&=0\,,&  \quad \partial_{r}\widehat \alpha^{22,22}_{(22)_n}[(11)_m]&=0 .\label{eq:gffduality}
	\ea
	
	Similar equations can be written for $\beta^{22,22}_{(22)_n}$, by moving the Kronecker delta to the right column.
	These conditions are understood as follows. The first line are the original conditions satisfied by the functionals. Adding the second line implies that the new functionals now have (double) zeros acting on the $(11)_m$ operators --- recall this means evaluating the functionals on the right dimensions and OPE orientation, and in this case we mean the $(11)_m$ operator for the $\phi,\phi^2$ system. Finally the last line guarantees that these vanishing conditions are still true under small deformations of the OPE orientation. Essentially the second and third lines ensure that the functional does not change sign in the neighbourhood of its zeros. While these two extra sets of   conditions are not necessary for diagonalizing the bootstrap equations for the mixed GFF solution, they do allow to have diagonal equations even slightly away from this solution.
	
	In practice the new duality conditions can be satisfied by setting:
	\begin{equation}
		\widehat \alpha^{22,22}_{(22)_n}=\alpha^{22,22}_{(22)_n}-\sum_{m=0}^\infty\left[c_n^m\alpha^{11,11}_{(11)_m}+d_n^m \alpha^{11,22}_{(11)_m} +e_n^m \beta^{11,11}_{(11)_m} \right],\label{eq:dressing}
	\end{equation}
	and tuning coefficients appropriately. The same procedure can be applied to $\beta^{22,22}_{(22)_n}$. 
	
	%
	%
	\vspace{0.5cm}

	\begin{figure}[t]
		\includegraphics[width=8.5 cm, height=7 cm]{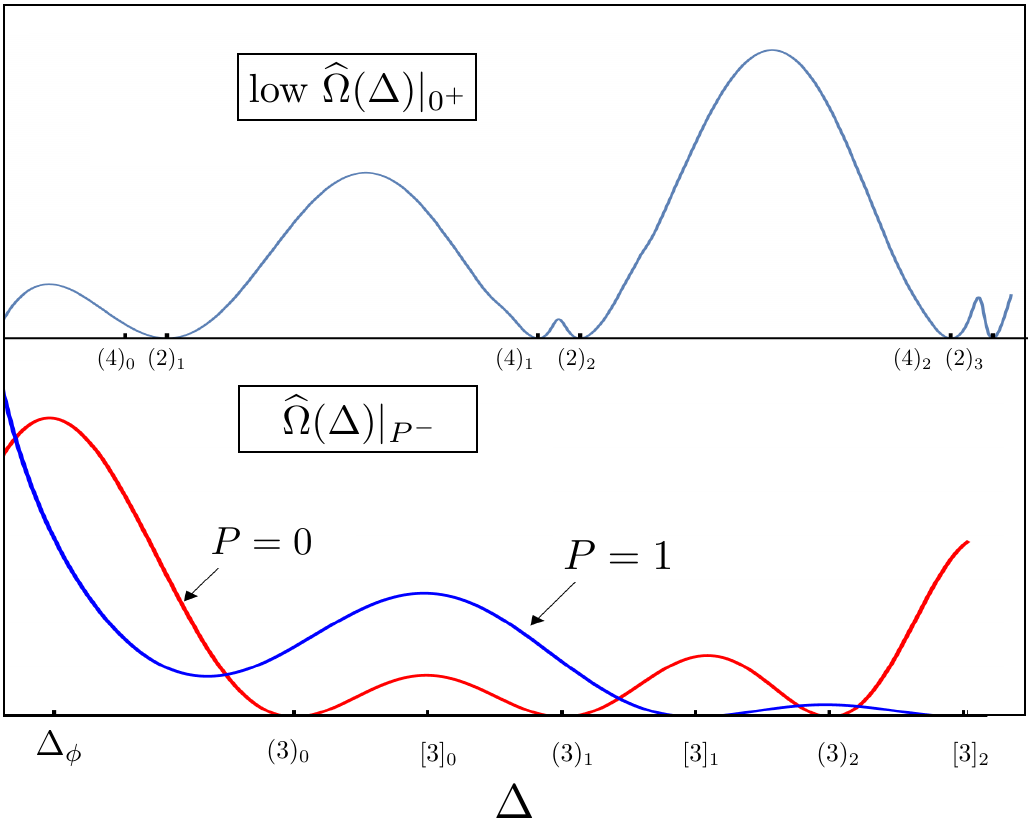}
		\caption{The $\widehat \Omega$ functional action for $\Df=9/10$  (height rescaled  for clarity; $(n)_k$ and $[n]_k$ denote $\Delta=n\Df+2k$ and $\Delta=n \Df+2k+1$ respectively). It  is a $2\times 2$ matrix corresponding to $11$ and $22$ OPE channels. Top ($Z_2$ even sector): We plot its lowest eigenvalue which is non-negative for $\Delta\geq 2\Df+2$. At zeros the corresponding eigenvector is proportional to $\{\lambda_{\cO}^{11},\lambda_{\cO}^{22}\}$ of the GFF solution. Bottom ($Z_2$ odd sector): We see positivity is achieved for any gap, but optimal bound requires compatibility with GFF.}
		\label{fig:low}
	\end{figure}
	
	{\em An optimal bound --- } We will now show that using a functional closely related to the above we can obtain an optimal bound satisfied by  GFF. Let us consider the same setup as before, with operators labeled by their $Z_2$ ($\pm$) and spacetime parity  ($0,1$ for even/odd) quantum numbers, e.g. $0^+$ etc. We set $\phi\equiv\Phi_1$ and  $\phi^2\equiv \Phi_2$ with dimensions and $Z_2$ charges as before. We assume that $\phi$ and $\phi^2$ are unique and also the leading non-identity operators, and fix the OPE data of the $\phi^2$ operator, (i.e. $a^{ij,kl}_{\phi^2}$) to the GFF values, which with the uniqueness assumption i.e. \reef{eq:nondeg} also fixes $a^{12,12}_\phi$. Our analytic bound is  more simply stated for $\Df\leq 1$, which we do here, leaving further discussion to \cite{Note1}.
	In this case our last assumption is that in the $0^+$ sector there should be no other operators below a gap no smaller than $\simeq 2\sqrt{2} \Df$, apart from $\phi^2$.

	Under these assumptions, consider a $0^+$ operator denoted $\phi^4$ and of dimension $4\Delta_\phi$. Then we claim there is an upper bound on $a^{22,22}_{\phi^4}$  which is saturated by the GFF solution. To prove this bound we first define $\Omega:=\alpha^{22,22}_{(22)_0}+ a \alpha^{11,11}_{(11)_0} + b\alpha^{12,12}_{[12]_0}+c \beta^{12,12}_{[12]_0}$. For any $a,b,c$ we can now ``dress'' this functional so that it is orthogonal to states in the 11 OPE. Specifically, we add to $\Omega$ an infinite sum as in \reef{eq:dressing} and
	impose most of the duality conditions \reef{eq:gffduality} except the one involving the state $(11)_0$ to obtain a new dressed functional $\widehat \Omega$. We find that for suitable choices of $a,b,c$ in some range $\widehat \Omega$ is positive semidefinite for scaling dimensions consistent with the gap assumptions above, see figure \ref{fig:low}. Thus it leads to a bound
	\ba\label{eq:an-bound}
	a^{22,22}_{\phi^4}\leq -\left(\widehat\Omega[\mbox{Id}]+a^{12,12}_{\phi}\widehat\Omega[\phi] +a^{22,22}_{\phi^2} \widehat\Omega[\phi^2]\right)=a^{22,22}_{\phi^4,\text{gff}}.
	\ea
	To understand the equality, note that by construction $ \widehat\Omega$ annihilates every operator appearing in the GFF solution, except for $\phi,\phi^2$ and $\phi^4$. This implies that for that solution the above inequality becomes an identity, i.e. the bound is saturated. To show this it is important to note that in the GFF solution there is no operator $[12]_0$ (it is $\partial \phi^3$, a descendant), otherwise $\alpha_{[12]_0}^{12,12}$ would be sensitive to it.

	To conclude, let us make a few comments on our assumptions. It may seem surprising that we need to impose a gap in the $Z_2$ even sector even after completely fixing the data of $\phi^2$. Indeed, a non-trivial but true fact is that fixing this data to its GFF values already implies that the {\em entire} 11 OPE must be that of a GFF~\cite{Mazac:2018ycv}. It is easy to see that this means that the 1111 and 1122 four-point functions are automatically the same as the GFF ones, and that the $(11)_n$ double traces appear in the 22 OPE with their GFF OPE values. However, at this point the 22 OPE is still not fully constrained. Maximizing the OPE of the operator $\phi^4$ fixes it to become that of the GFF solution, but only under our gap assumption in the $0^+$ sector -- OPE bounds generally require a minimal gap, otherwise the maximum is infinite.
	
	\subsubsection{Numerical explorations: islands with GFF inhabitants and particle production}
	\label{app:island}

\begin{figure}[t]
		\centering
		\includegraphics[width=7.5 cm,height=6 cm]{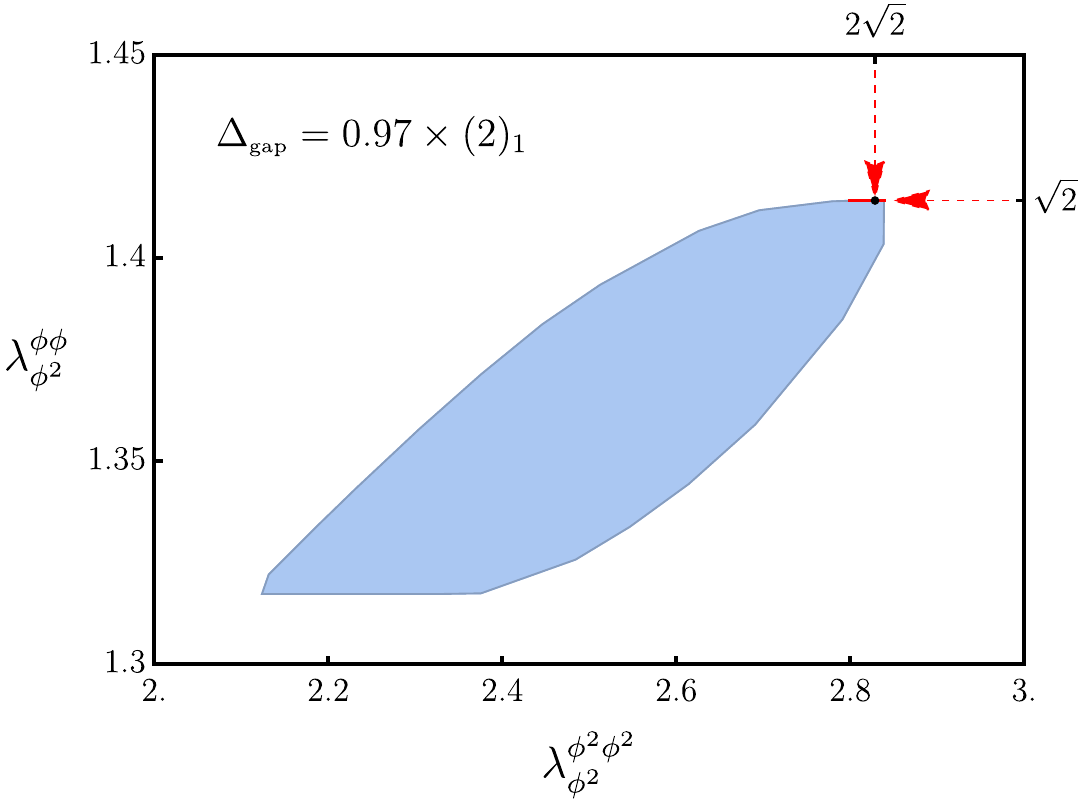}
		\caption{ Feasible region in $(\lambda^{\phi^2\phi^2}_{\phi^2},\lambda^{\phi\phi}_{\phi^2})$ for $\Delta_{\phi}=\frac{7}{3}$ and a gap in the $0^+$ sector equal to $\sim 6.475$. When the gap is set to $2\Df+2$ the region shrinks to a small line segment indicated in red whose rightmost tip seems to tend to the GFF solution.
		}
		\label{fig:phi2}
	\end{figure}

	In this section we perform some preliminary numerical explorations using our setup \footnote{We used SDPB\cite{Simmons-Duffin:2015qma} software to solve the optimization problem involving mixed correlators.}, leaving more detailed computations for future work. In figure \ref{fig:phi2} we explore again the same mixed correlator system as before looking
	for the feasible region in the $\lambda^{11}_2,\lambda^{22}_2$  plane i.e. the OPE data of the operator $\phi^2$ \footnote{In \cite{Antunes:2021abs} this was also explored using traditional bootstrap methods.}.
	
	The island shown appears when we set gaps in the $Z_2$ odd sectors. In particular in the figure we have set these to be $3\Delta_{\phi} \ (3\Delta_{\phi}+1)$ in parity even (odd) channels respectively. When the gap is smaller, say $\Delta_{\phi}$ in $0^-$, we observe a region where the left direction of the plot is qualitatively similar, while on the right, it extends to a strip-like shape without any upper bound for $\lambda^{\phi^2\phi^2}_{\phi^2}$, as found in \cite{Antunes:2021abs}. Conversely, larger gaps further shrink the allowed region.
	The island displays two seemingly flat regions in the bottom and top, which are not numerical artifacts. In fact they are well described by single correlator bounds on $\lambda^{\phi\phi}_{\phi^2}$. This suggests that the 11 OPE is unchanged along these boundaries. On the top, the
	bound is simply $\sqrt{2}$ i.e. the GFF value. When we examine the spectrum of the extremal solution corresponding to the top right corner of the feasible region, we find a solution consistent with a non-interacting (GFF) 11 and 12 OPEs and an interacting 22, see figure 1 in the supplementary material. As explained there, this can be thought of as a consequence of the existence of a $\Phi^8$ deformation in AdS$_2$ which allows to deform the 22 OPE while holding the 11 fixed.
	While in this special case the 11 OPE remains free, the general result is that it should not only become interacting but also manifest ``particle production'', i.e. couplings to the $(22)_n$ operators. An example is shown in Fig. 2 of the supplementary material.

	\subsubsection*{\label{sec:disc}Discussion and outlook}
	In this paper we have introduced a presumably complete reformulation of crossing symmetry constraints for an arbitrary set of 1d CFT correlators. Our sum rules are naturally adapted to study deformations of generalized free fields, allowing us to bridge the gap between analytic and numeric computations and promising to help us pinpoint desired theories to bootstrap. Our initial numerical explorations, which will be developed elsewhere, show not only greatly improved speed of convergence relative to traditional bootstrap methods, but much greater accuracy, allowing us to give precise spectra for interacting CFT solutions with features like ``particle production'' or lack thereof. Our methods thus promise to push the bootstrap into much larger sets of mixed correlators and probe the high energy asymptotics of CFT correlators. We will return to these questions in the very near future.

	%
	
	\begin{acknowledgments}
		We are grateful for discussions with Romain Usciati, Nat Levine, Antonio Antunes, Zechuan Zheng. A.K. is supported by DFG (EXC 2121: Quantum Universe, project 390833306). This work was co-funded by the European Union (ERC, FUNBOOTS, project number 101043588). Views and opinions expressed are however those of the author(s) only and do not necessarily reflect those of the European Union or the European Research Council. Neither the European Union nor the granting authority can be held responsible for them. 
		
	\end{acknowledgments}

\bibliography{mybib}
	
	\vspace{2cm}

\pagebreak
\widetext
 \appendix

	\section{Witten diagrams and functional actions}
	\label{app:witten}
	Here we will elaborate on the  Witten diagrams in AdS$_2$ and evaluating their conformal block expansions. 
	In this work, we introduced two different kinds of Polyakov blocks $\mathcal{P}_{\mathcal{O}}^{ij,kl}$ corresponding to even or odd parity of the operator $\cO$. They are given as sums of scalar  or spin 1  exchange Witten diagrams respectively along with relevant Regge bounded contact diagrams. Below our strategy will be to compute these objects for general spacetime dimension and set $d=1$ at the end of the computation. Note that a general $d$ Witten diagram will depend on the  spin $\ell$ of the exchange operator while in $d=1$ the $\ell$ dependence will be replaced with parity $P$. \\
	
	{\em Expanding Witten diagrams in conformal blocks --- } For general spacetime dimension $d$ and scaling dimension and spin $\Delta,\ell$ of an exchanged operator, $s$-channel Witten diagrams are defined as  \cite{DHoker:1999mqo},
	\begin{equation}
		\begin{split}
			W^{ij,kl,s}_{\Delta,\ell}(x_i)=& \int \frac{d^{d+1}z_1}{z_{10}^{d+1}}\frac{d^{d+1}z_2}{z_{20}^{d+1}} K_{\Delta_i}(z_1,x_1)(\nabla^\mu)^{\ell} K_{\Delta_j}(z_2,x_2)\\
			& \pi^{\Delta}_{\mu_1..\mu_{\ell_e}\nu_1..\nu_{\ell} }(z_1,z_2) K_{\Delta_k}(z_2,x_3)(\nabla^\nu)^{\ell} K_{\Delta_l}(z_2,x_4),
		\end{split}
	\end{equation}
	where $\pi^{\Delta}_{\mu \nu}(z_1,z_2)$ and $K_{\Delta_i}(z_1,x_1)$ are suitably normalized bulk to bulk and bulk to boundary propagators respectively. The corresponding crossed channel exchanged Witten diagram is obtained by swapping $x_i$ and the corresponding dimension of the external operator. So the $t$-channel corresponds to the change of labels $j\leftrightarrow l$ while the $u$-channel corresponds to $j\leftrightarrow k$. 
	
	We want to compute conformal block decompositions of the 1d Witten diagrams. So we focus on only $\ell=0,1$ in the bulk which corresponds to $P=+,-$ respectively.
	They can be written in terms of a single cross-ratio  \footnote{In general spacetime dimension these quantities depend on two cross ratios $z,\bar{z}$.} as follows
	\be
	W^{ij,kl,c}_{\Delta,P}(x_i)= \frac{1}{|x_{12}|^{\Delta_1+\Delta_2}|x_{34}|^{\Delta_3+\Delta_4}}\left(\frac{x_{14}^2}{x_{24}^2}\right)^{\frac{\Delta_{12}}{2}}\left(\frac{x_{14}^2}{x_{23}^2}\right)^{\frac{\Delta_{34}}{2}}W^{ij,kl,c}_{\Delta,P}(z)
	\ee
	Now this allows a conformal block decomposition that can be written as (denoting $E=ij,kl$) 
	\begin{align} 
		W^{E,s}_{\Delta,P}(z)&=G_{\Delta}^E(z)-\sum_{I_s\in\mathcal I_{s}^{E}} \alpha_{I_s}^{E,s}(\Delta,P) G^E_{\Delta_{I_s}}(z)\,.\label{wittendecomp}\\
		W^{E,t}_{\Delta,P}(z)&=-\sum_{I_s\in\mathcal I_{s}^{E}} \alpha_{I_s}^{E,t}(\Delta,P) G^E_{\Delta_{I_s}}(z)\,.\label{wittendecomp1}\\
		W^{E,u}_{\Delta,P}(z)&=-\sum_{I_s\in\mathcal I_{s}^{E}} \alpha_{I_s}^{E,u}(\Delta,P) G^E_{\Delta_{I_s}}(z)\label{wittendecomp2}
	\end{align}
	Here $\mathcal I_{s}^{E}$ refers to the set of double trace operators $(ij)_n,[ij]_n,(kl)_n,[kl]_n$. 
	Also $G_{\Delta}^{E}(z)$ is the 1d conformal block given by
	\ba
	G_{\Delta}^{ij,kl}(z)=\frac{z^{\Delta}}{z^{\Delta_i+\Delta_j}}\, _2F_1(\Delta-\Delta_{ij},\Delta+\Delta_{kl},2\Delta,z)\,.\label{eq:1dblock}
	\ea
	The coefficients $\alpha_{I_s}^{E,c}$ can be obtained by using the Casimir operator to relate exchange Witten diagrams to contact diagrams as follows \cite{Zhou:2018sfz,DHoker:1999mqo} 
	\be\label{casWitten}
	(\mathcal{C}_{12}+c_{\Delta,P})W^{E,s}_{\Delta,P}(z)=N^E_{\Delta,P}W^{E, \text{con}}(z).
	\ee
	Here $\mathcal{C}_{12}$ is the conformal Casimir operator acting on $x_1$ and $x_2$ \cite{Dolan:2003hv} and $c_{\Delta,P}$ the corresponding eigenvalue for an operator with dimension $\Delta$ and parity $P$. In the righthand side $W^{E, \text{con}}$ is a combination of  tree level contact diagrams which involve four scalar fields with a 4-point interaction vertex. Also $N^E_{\Delta,P}$ is a normalization constant. Given that the maximum spin for the exchanged bulk operators is 1 for our case, the above relation leads to contact diagrams involving vertices with derivatives up to a maximum of second order. There are similar identities for any spin $\ell$ exchange Witten diagram in higher dimensions.
	
	It is very easy to work out the block decomposition of any contact diagram  from their simple form in Mellin space which we discuss below. Notice the Casimir operator acts on the $s$-channel conformal block diagonally since blocks are eigenfunctions of Casimir. By applying the operator to the expression \eqref{wittendecomp}, we can extract the values of $\alpha^{E,s}_{I_s}$ directly in terms of the conformal block decomposition of contact diagrams. 
	
	To obtain $\alpha^{E,t}_{I_s}$ or $\alpha^{E,u}_{I_s}$ i.e. the  decomposition of the crossed channel exchange Witten diagram in the $s$-channel conformal blocks, we employ a similar approach with the appropriate changes in the variables $x_i$ in \eqref{casWitten}. However, it is important to note that in the crossed channel, the action of the crossed channel Casimir on the direct channel conformal block is not diagonal. As a result, instead of obtaining a closed-form expression, we encounter a recursion relation for the coefficients of the decomposition \cite{Zhou:2018sfz}. This is very easy to implement numerically provided the seed block, i.e., the leading block decomposition coefficient $\alpha^{E,c}_{(ij)_0}$ or $\alpha^{E,c}_{(kl)_0}$ (for $c=t,u$). This we can obtain from the Mellin amplitude representation for the Witten diagrams \cite{Mack:2009mi,Penedones:2010ue} which we describe below. \\
	
	{\em Witten diagrams in Mellin space --- } The Mellin amplitude of a general CFT correlation function can be defined by the inverse Mellin transform
	\cite{Gopakumar:2016cpb}:
	\begin{equation}\label{Mellindef}
		\mathcal G(u,v)=\int [ds][dt] u^s v^t \Gamma(s+t+a)\Gamma(s+t+b)\Gamma(-t)\Gamma(-t-a-b)\Gamma(\frac{\Delta_i+\Delta_j}{2}-s)\Gamma(\frac{\Delta_k+\Delta_l}{2}-s) M(s,t),
	\end{equation}
	where, $a=\frac{\Delta_j-\Delta_i}{2}$,  $b=\frac{\Delta_k-\Delta_l}{2}$. Here $u,v$ are the conformal cross-ratios in general $d$. 
	The Mellin amplitude of $s$-channel spin-$\ell$ exchange Witten diagram $W^{E,s}_{\Delta,\ell}$ is given by
	\begin{equation}
		\begin{split}
			& M^{E,s}_{\Delta,\ell}(s,t)= \frac{2}{(\Delta-\ell-2s)\Gamma(1-\frac d2+\Delta)}\Gamma\left(-\frac d2+\frac{\ell}{2}+\frac{\Delta}{2}+\frac{\Delta_i}{2}+\frac{\Delta_j}{2}\right)\Gamma\left(-\frac d2+\frac{\ell}{2}+\frac{\Delta}{2}+\frac{\Delta_k}{2}+\frac{\Delta_l}{2}\right) \\
			& _3F_2\Bigg[\left\{\frac{\Delta-\ell}{2}-s,1+\frac{\Delta-\ell}{2}-\frac{\Delta_i+\Delta_j}{2},1+\frac{\Delta-\ell}{2}-\frac{\Delta_k+\Delta_l}{2}\right\},\left\{1+\frac{\Delta-\ell}{2}-s,1-\frac d2+\Delta\right\},1\Bigg]\\
			& \frac{P_{\Delta-\frac d2,\ell} (s,t)}{(d-\Delta-1)_\ell(\Delta-1)_\ell}.
		\end{split}
	\end{equation}
	where the Mack polynomial is \cite{Mack:2009mi,Mack:2009gy},
	\begin{equation} 
		\begin{split}
			&   P_{\Delta-\frac d2,\ell}(s,t)=\sum_{m+n=0}^{\ell}\frac{(-1)^{m+n} \ell!}{n! (\ell-m-n)!} \frac{\Gamma\left(\frac{-2+d+2\ell}2\right)\Gamma\left(\frac{-2+2a+2b+d+2m+2n}{2}\right)}{\Gamma\left(\frac{d-2}2\right) \Gamma(m+1)\Gamma\left(\frac{2a+2b-2+d+2n}2\right)\Gamma(\ell-1+2\bar{\lambda}_2) }\\
			& _4F_3\bigg[\{-m,3-d-\ell-n,1-a-\bar{\lambda}_1,1-b-\bar{\lambda}_1\},\{2-\frac d2-\ell,2-a-b-\frac d2-m-n,2-2\ell-2\bar{\lambda}_2\},1 \bigg]\\
			&\frac{\Gamma(-1+2\ell+2\bar{\lambda}_2)}{(d-2)_\ell} (a+m-\ell+n+\lambda_1)_{\ell-m-n} (b+m-\ell+n+\lambda_1)_{\ell-m-n} (-1+\ell+2\lambda_2)_n (\lambda_2-s)_m (-t)_n,
		\end{split}
	\end{equation}
	and
	\ba
	\lambda_1=\frac{\Delta+\ell}{2},\quad \,\bar{\lambda}_1=\frac{d-\Delta+\ell}{2},\quad\, \lambda_2=\frac{\Delta-\ell}{2},\quad\, \bar{\lambda}_2=\frac{d-\Delta-\ell}{2}.
	\ea
	The $t$-channel exchange Witten diagram is given by the following change of variables,
	\begin{equation}
		s \xrightarrow[]{} t+\frac{1}{2}(\Delta_j+\Delta_k),\,\,\,\,\,\,\,\,\,\,\,\, t\xrightarrow[]{} s-\frac{1}{2}(\Delta_k+\Delta_l),
		\,\,\,\,\,\,\,\,\,\,\,\,\Delta_j \leftrightarrow \Delta_l.
	\end{equation}
	Similarly the $u$-channel exchange Witten diagram is given by the following change of variables,
	\begin{equation}
		s \xrightarrow[]{} \frac{1}{2}(\Delta_i+\Delta_l)-s-t,\,\,\,\,\,\,\,\,\,\,\,\, t\xrightarrow[]{} t,\,\,\,\,\,\,\,\,\,\,\,\,\Delta_j \leftrightarrow \Delta_k.
	\end{equation}
	The Mellin ampllitude of the Regge-bounded contact diagrams is of the form $ps+qt+r$ ($p,q,r$ are constants).

	Let us now restrict to $d=1$ again. For this we will set  $u=z^2, v=(1-z)^2$ in \eqref{Mellindef}. 
	The seed conformal block that enters the recursion relation for crossed channel exchange Witten diagrams corresponds to the poles at $s=\frac 12(\Delta_i+\Delta_j)$ or $s=\frac 12(\Delta_k+\Delta_l)$.  The respective coefficient is obtained by computing the corresponding residue and then performing the $t$ integration.  The result can be written in terms of a $_7F_6$ hypergeometric function \cite{Gopakumar:2018xqi}.  \\
	
	{\em Special case: integer differences in external dimensions --- } We end this appendix by commenting on the special case when $\Delta_i+\Delta_j$ differs from $\Delta_k+\Delta_l$ by an even integer. In this situation there is a pole in the conformal block decomposition coefficients $\alpha^{E,c}_{I_s}(\Delta)$. This comes from the factor like $\Gamma(\frac{\Delta_i+\Delta_j}{2}-s)$ in the Mellin amplitude definition \eqref{Mellindef}. Let us consider the following expansion explicitly 
	\begin{equation}\label{CBdecodemo}
		W^{E,c}_{\Delta,+}(z)=\sum_n \alpha^c_{(ij)_n}G^{ij,kl}_{(ij)_n}(z)+\sum_n \alpha^c_{(kl)_n}G^{ij,kl}_{(kl)_n}(z)+\cdots \,.
	\end{equation}
	Here the ellipsis denotes other families of blocks appearing in the decomposition. When $\Delta_i+\Delta_j \to \Delta_k+\Delta_l$ the expansion of the  block decomposition coefficients near the pole looks like,
	\begin{equation}
		\begin{split}
			&  \alpha^{E,c}_{(ij)_n}\rightarrow \frac{\beta_{(ij)_n}^{E,c}}{(\Delta_i+\Delta_j)-(\Delta_k+\Delta_l)}+\omega^{E,c}_{(ij)_n}\\
			& \alpha^{E,c}_{(kl)_n}\rightarrow \frac{\beta_{(kl)_n}^{E,c}}{(\Delta_i+\Delta_j)-(\Delta_k+\Delta_l)}+\tilde{\omega}^{E,c}_{(ij)_n}
		\end{split}
	\end{equation}
	We will find that $\beta_{(ij)_n}^{E,c}=-\beta_{( kl)_n}^{E,c}$. This will give rise to derivative of blocks (denoting $\partial G^E_{\Delta_0}:=(\partial_{\Delta}G^E_{\Delta})_{\Delta=\Delta_0}$) appearing in the expansion \eqref{CBdecodemo}
	\begin{equation}
		W^{E,c}_{\Delta,+}(z)=\sum_n \alpha^{E,c}_{(ij)_n}G^{ij,kl}_{(ij)_n}(z)+\sum_n \beta^{E,c}_{(ij)_n}\partial G^{ij,kl}_{(ij)_n}(z)+\cdots .
	\end{equation}
	We also abuse the notation $\alpha^c_{(ij)_n}$ which now denotes the other piece $\big(\omega^{E,c}_{(ij)_n}+\tilde{\omega}^{E,c}_{(ij)_n}\big)$. In other words, we get
	\ba
	\{\alpha^{E,c}_{(ij)_n},\alpha^{E,c}_{(kl)_n}\}&\to \{\alpha^{E,c}_{(ij)_n},\beta^{E,c}_{(ij)_n}\},\\
	\{\alpha^{E,c}_{[ij]_n},\alpha^{E,c}_{(kl)_n}\}&\to \{\alpha^{E,c}_{[ij]_n},\beta^{E,c}_{[ij]_n}\}.
	\ea

	\section{Checks with analytical solutions}
	
	In section 2 of our main text we pointed out that the sum rules eq. (13) of main text trivialize on a set of correlators in a theory of decoupled GFF scalars. We also showed how the sum rules can bootstrap the system of $\phi$ and $\phi^2$ fields in GFF.  In this appendix we will address the checks on  both of these theories more elaborately and clarify  the steps involved. For both these examples our general argument will be as follows: first we consider a set of mixed correlators that satisfy the crossing symmetry relation:
	\bea
	\mathcal G^{ij,kl}(z)=\left(\frac{1-z}{z}\right)^{\Delta_{i}+\Delta_j-\Delta_k-\Delta_l}\mathcal G^{il,kj}(1-z).\label{eq:crossing}
	\eea
	Then we will show how the sum rules solve the entire set of OPE data that is consistent with these correlators. \\

	{\em Tensor product theories --- } Our first test is that the sum rules correctly bootstrap correlators in a tensor product theory of several GFF operators. Thus we will set $\Phi_i\equiv \phi_i,$ a fundamental generalized free field, and $\phi_i,\phi_j$ do not interact for $i\neq j$.
	Let us explain how to bootstrap correlators in the tensor product theory of two bosonic fields $\phi_1, \phi_2$ of positive parity. They are given as
	\ba
	\mathcal G^{ii,ii}(z)&=\frac{1}{z^{2\Delta_i}}+\frac{1}{(1-z)^{2\Delta_i}}+1\equiv \mathcal G_{\Delta_i}^{\mbox{\tiny gff}}(z)\\
	\mathcal G^{12,12}(z)&=1\,, \qquad%
	\mathcal G^{11,22}(z)=\frac{1}{z^{2\Delta_1}}\,.\label{eq:gff}
	\ea
	Consider first the sum rules for $E=11,11$. In this case the functional actions become
	\ba
	\omega^{1111}_{(11)_n}[\cO]:=(r^{11}_{(11)_n})^2 \omega_n^{\Delta_1}(\Delta_{\cO})\,,
	\ea
	with $\omega=\alpha,\beta$ and
	\begin{equation}
		\omega_n^{\Delta_1}(\Delta_{\cO}):=\left[\omega^{11,11,s}_{(11)_n}(\Delta_\cO,+)+\omega^{11,11,t}_{(11)_n}(\Delta_\cO,+)+\omega^{11,11,u}_{(11)_n}(\Delta_\cO,+)\right].
	\end{equation} 
	The $\omega_n^{\Delta_i}(\Delta)$ with $\omega=\alpha,\beta$ are nothing but the functionals already introduced in \cite{Mazac:2016qev,Mazac:2018mdx,Mazac:2018ycv}, so in this case our sum rules reproduce the ones derived there. More interestingly, consider the mixed correlators 12,12 and 11,22. The sum rules now are
	\ba
	\sum_{\cO} \lambda^2_{\cO} \,\omega[\cO]=0\,,\ \omega\in\left\{ 
	\begin{array}{c}
		\alpha^{12,12}_{(12)_n},
		\beta^{12,12}_{(12)_n},
		\alpha^{12,12}_{[12]_n},\vspace{0.2cm}\\
		\beta^{12,12}_{[12]_n},
		\alpha^{11,22}_{(11)_n},\alpha^{11,22}_{(22)_n}
	\end{array}
	\right\}
	\ea
	with $n=0,1,\ldots$ However, in this case there are two relevant contact diagrams possible, allowing us to make two subtractions. We choose them to eliminate the functionals $\beta^{12,12}_{(12)_0}$ and $\beta^{12,12}_{[12]_0}$. With this caveat, let us check that the tensor product theory satisfies the sum rules. The operators appearing above are the identity, as well as double traces $(11)_n,(22)_n,(12)_n$ and $[12]_n$. Furthermore each such operator always appears in a single OPE channel, which means the corresponding $r_\cO$ vector is essentially a Kronecker delta. We have then e.g.
	\ba
	\alpha^{12,12}_{(12)_n}[\text{Id}]+\sum_c \sum_{I_c} \lambda^2_{I_c} \alpha_{(12)_n}^{12,12}[I_c]=0 
	\ \Leftrightarrow \ (\lambda^{12}_{(12)_n})^2=-\alpha^{12,12,u}_{(12)_n}(0,+)\,.
	\ea
	Similarly we find $(\lambda^{12}_{[12]_n})^2=-\alpha^{12,12,u}_{[12]_n}(0,-)$. We have extensively checked that these results match those extracted from the actual CFT correlators. For $E=11,22$ we get
	\ba
	(\lambda^{11}_{(ii)_n} \lambda^{22}_{(ii)_n})&=-\alpha^{11,22}_{(ii)_n}(0).
	\ea
	The righthand side is actually zero, yielding the expected result $\lambda^{ii}_{(jj)_n}=0$ for $i\neq j$ in the tensor product theory. Finally, the $\beta$ sum rules are satisfied only if ~$\beta^{12,12}_{(12)_n}[\text{Id}]=\beta^{12,12}_{[12]_n}[\text{Id}]=0$, which is indeed the case. 
	\vspace{0.5cm}
	
	{\em The $\phi,\phi^2$ system --- }
	The main application in section 2 of main text  is to bootstrap a set of crossing symmetric CFT correlators built from the $\phi$ and $\phi^2$ GFF operators. This means we have two operators of dimension $\Delta_1=\Delta_\phi$ and $\Delta_2=2\Delta_1$ whose correlators satisfy eq. (17) of the main text, which we repeat for convenience:
	\ba
	\mathcal G^{11,11}(z)&=\mathcal G_{\Df}^{\mbox{\tiny gff}}(z)\\
	\mathcal G^{12,12}(z)&=1+a^{12,12}_1\,\left[\frac{1}{z^{2\Df}}+\frac{1}{(1-z)^{2\Df}}\right]\\
	\mathcal G^{11,22}(z)&=\frac{1}{z^{2\Df}}+a^{12,12}_1\left[1+\frac{1}{(1-z)^{2\Df}}\right]\\
	\mathcal G^{22,22}(z)&=\mathcal G^{\mbox{\tiny gff}}_{2\Df}(z)+
	+\frac{a^{22,22}_2}2
	\frac {z^{2\Df}+(1-z)^{2\Df}+1}{z^{2\Df}(1-z)^{2\Df}}\,. 
	\label{eq:gfffamily1}
	\ea
	There are two free parameters in these expressions, with different choices corresponding to different physical theories, so the sum rules will be unable to fix them. For instance we could set $\Phi_1:=\phi_1$ and $\Phi_2:=(2N)^{-\frac 12} \sum_{k=1}^N \phi_k \phi_k$ in a theory of $N$ decoupled generalized free fields, and then $a_2^{22,22}=8/N$. Setting  $a_1^{12,12}=a_2^{22,22}=0$ corresponds to $N\to \infty$ and $\Phi_1, \Phi_2$ become two decoupled GFF fields as in our previous example. Below we will focus on the case where $\Phi_1=\phi$ and $\Phi_2=\phi^2$, with the exchanged field in the $12$ OPE identified with $\phi$, and with $\phi^2\equiv (11)_0$. These  correspond to imposing the two conditions:
	\ba
	a_1^{12,12}=a_{(11)_0}^{11,11}\,, \quad a_2^{22,22}=(a_{(11)_0}^{11,22})^2/a_{(11)_0}^{11,11}\,. \label{eq:nondeg1}
	\ea
	To bootstrap these correlators we write down the most general OPE consistent with a $Z_2$ symmetry and the double trace operator spectrum. Schematically:
	\ba
	\phi\times \phi& =\text{Id}+\sum_n (11)_n+\sum_n (22)_n,\\
	\phi\times \phi^2 &=\phi+\sum_n (12)_n+\sum_n [12]_n,\\
	\phi^2\times \phi^2 & =\text{Id}+\sum_n (11)_n+\sum_n (22)_n\,.
	\ea
	We begin by considering the mixed correlator constraints. The states in the 12 OPE are determined by:
	\ba
	a^{12,12}_{(12)_n}=-\alpha^{12,12,u}_{(12)_n}(0,+)-a^{12,12}_1 \alpha^{12,12,s}_{(12)_n}(\Delta_1,+)
	\ea
	and similarly for the parity odd ones. There are also $\beta^{12,12}$ equations. These turn out to be non-trivially satisfied thanks to the condition $\Delta_2=2\Delta_1$. Moving on, we also have equations:
	\begin{equation}
		a^{11,22}_{(11)_n}=-\underbrace{\alpha^{11,22,s}_{(11)_n}(0,+)}_{=0}\\-a^{12,12}_1\left[\alpha^{11,22,t}_{(11)_n}(\Delta_1,+)+\alpha^{11,22,u}_{(11)_n}(\Delta_1,+)\right]\label{eq:11n}
	\end{equation}
	and the righthand side is non-trivial. In particular one finds
	\ba
	a^{11,22}_2\equiv a^{11,22}_{(11)_0}=a_1^{12,12}
	\ea
	consistently with the expected solution.
	For $a^{11,22}_{(22)_n}$ we get an equation analogous to \reef{eq:11n}, but now non-trivially the whole righthand side turns out to be zero, where again the condition $\Delta_2=2\Delta_1$ is crucial. So we learn that as expected the $(22)_n$ states do not appear in the $11$ OPE. This means that the $11,11$ correlator contains only $(11)_n$ and is now easily bootstrapped as a fundamental GFF correlator: 
	\ba
	a^{11,11}_{(11)_n}=-\alpha^{\Delta_1}_{n}(0)\equiv a_{\Df,n}^{\text{gff}}\,.
	\ea
	Finally, this leaves us the equations for the $22,22$ correlator. These are
	\ba
	a^{22,22}_{(22)_n}=\underbrace{-\alpha_n^{\Delta_2}(0)}_{=a^{\text{gff}}_{2\Df,n}}-\sum_{m=0}^\infty a^{22,22}_{(11)_m} \alpha_n^{\Delta_2}(\Delta_{(11)_m}),\\
	0=-\underbrace{\beta_n^{\Delta_2}(0)}_{=0}-\sum_{m=0}^\infty a^{22,22}_{(11)_m} \beta_n^{\Delta_2}(\Delta_{(11)_m}).
	\ea
	Imposing that the double trace spectrum contains non-degeneracies is the statement:
	\ba
	a^{22,22}_{(11)_n}=(a^{11,22}_{(11)_n})^2/a^{11,11}_{(11)_n}. \label{eq:a222211}
	\ea
	One can check that the $\beta$ equations are solved by this ansatz. In fact, turning the logic around, we should really think that they {\em imply} \reef{eq:a222211}, so that these are not independent inputs. Finally the $\alpha$ equations determine the coefficients $a^{22,22}_{(22)_n}$. 
	\section{Further details on optimal and numerical bounds} 
	\label{boundhigherdfi}
	
	In this appendix we will give further details on the analytic and numerical bounds discussed in the main text.\\
	
	{\em Optimal bound saturated by GFF --- }
	Let us first discuss how the optimisation problem described in the main text must to be modified so as to be saturated by a GFF when $\Df\geq 1$. As mentioned there, our assumptions include the existence of an operator $\phi^2$ in the 11 OPE with the correct GFF OPE coefficient, and this in fact forces the entire 11 OPE to be that of a GFF. This, together with crossing for the 1122 correlator then implies that all $(11)_m$ operators also appear in the 22 OPE with their correct GFF coefficients. Thus at this point we can focus on understanding crossing for the 2222 correlator, keeping in mind the necessary presence of the identity and the tower of $(11)_m$ operators. The difficulty now is that from the point of view of crossing constraints for this correlator, on the one hand we know that we must demand a gap in the OPE of about $2\sqrt{2}\Df$ otherwise there can be no upper bounds on {\em any} OPE coefficients; effectively this is because below this bound we can always add unitary solutions to crossing without identity to the correlator which can drive up OPE coefficients to arbitrarily large values. On the other, hand as $\Df$ increases, we know from our previous assumptions that there will definitely need to be some of the $(11)_m$ operators appearing below this gap. We must therefore allow at least those operators to appear, otherwise we would have inconsistent assumptions. To resolve this tension, the solution is to impose a general gap in the $0^+$ sector of about $2\sqrt{2}\Df$, but still allow a finite number of operators with dimension $2\Df+2m$ below it. Furthermore we must demand that any such operators are {\em non-degenerate}. In this way, only the operators which are forced by the 11 OPE constraints will appear and no others (which could appear just in the 22 OPE). In practice imposing these non-degeneracy conditions must be done by scanning in the OPE orientation space of these operators. The claim then is that our functional establishes that in this space there is a single feasible point precisely when these orientations match the ones of the GFF solution, and at that point there is a bound on $\phi^4$ given by the full GFF mixed correlator system.\\
	
	{\em Extremal spectra and particle production --- }
	\begin{figure}[h]
		\centering
		\includegraphics[width=10 cm]{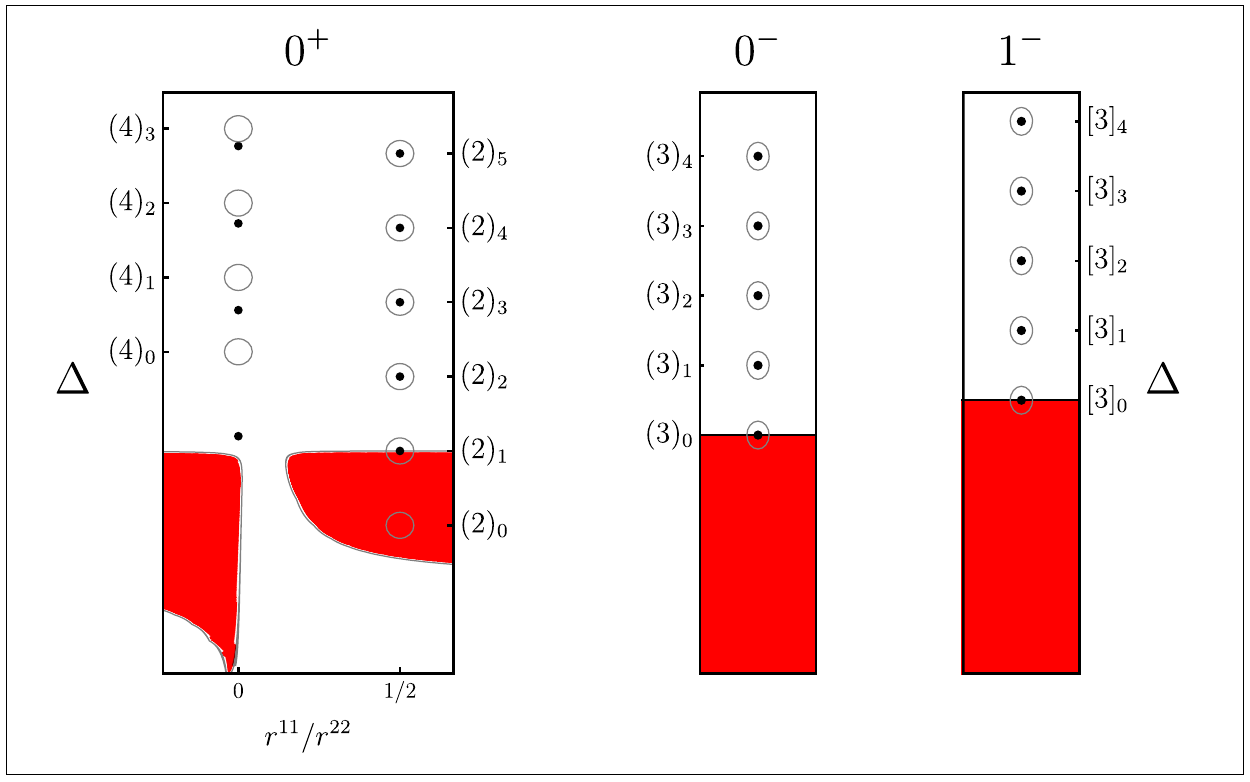}
		\caption{
			Functional positivity and extremal spectrum extracted from the top right corner of the feasible region of  Fig. 2 of main text. Red/white indicates negative/positive functional action. The black dots indicate zeros of the functional and black circles are centered around the GFF operators, with the labels $(n)_k$ and $[n]_k$ representing $\Delta=n\Df+2k$ and $\Delta=n \Df+2k+1$.
			In the $0^+$ sector this is a two dimensional plot: the functional at fixed $\Delta$ is a $2\times 2$ matrix which is then contracted with an OPE vector $(r^{11},r^{22})$. For other sectors the width in $x$ is for visual clarity.}
		\label{fig:phi2spec}
	\end{figure}
	In the main text we showed an island in the space of OPE coefficients for the $\phi^2$ operator in a mixed correlator system. The top right corner of the island corresponds to setting the $\phi^2$ OPE data to match that of GFF. As discussed above we expect that in this case the $1111$ and $1122$ correlators to match the GFF result, but not necessarily that of $22$. In figure \ref{fig:phi2spec} we examine the extremal spectrum corresponding to that corner, as obtained by looking for an upper bound on $\lambda^{\phi \phi}_{\phi^2}$ given $\lambda^{\phi^2 \phi^2}_{\phi^2}$. This spectrum is in perfect agreement with our expectations: the 11 OPE is indeed the GFF one, but we see that the 22 OPE  is not. To see this had to be the case, simply note that if it the 22 OPE contained the GFF operators then the resulting extremal functional would need to have double zeros on all $(22)_m$ operators. But such a functional would rule out the existence of the $\Phi^8$ bulk term, which can perturbatively modify the $(22)_m$ operators while keeping the $(11)_m$ unchanged.

\begin{figure}[h]
		\centering
		\includegraphics[width=10 cm,height=8 cm]{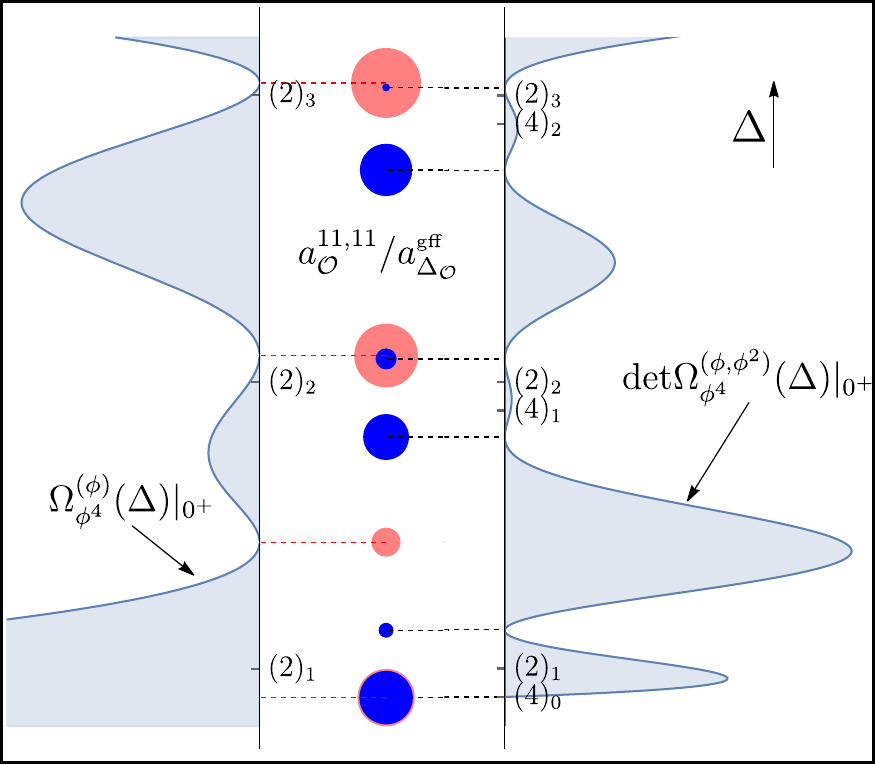}
		\caption{Interacting spectra and particle production.  Above we show results for $\Df=9/10$. On the left the extremal functional and resulting spectrum when including constraints from a single correlator. On the right, the determinant of the functional in the $0^+$ sector when accounting for a mixed correlator system. The notation $(k)_n$ represents $\Delta=k\Df+2n$. The disks in the center represent the size of the OPE coefficients as measured with respect to the GFF OPE density with external dimension $\Df$. We see that the mixed correlator system leads to a small decrease in the maximal value of $\lambda^{\phi\phi}_{\phi^4}$ and shifts in the dimensions of double trace operators in the $11$ OPE which decrease for large $\Delta$. However, it also forces 22 states to appear in the 11 OPE and a large redistribution in the OPE coefficients.   
		}
		\label{fig:production}
	\end{figure}

	Let us end this appendix with a small discussion on particle production. By particle production we mean   interacting solutions where an OPE channel, say 11, starts seeing new families of operators, e.g. both $(11)_n$ and $(22)_n$. We show such a solution in Fig. \ref{fig:production}. It follows from an optimization problem (for $\Df=9/10$) to maximize the OPE coefficient of an operator $\phi^4$ with dimension $4\Df$ in the presence of another operator $\phi^2$ in the OPE, with dimension $2\Df$ and OPE coefficient $\lambda^{\phi\phi}_{\phi^2}=\sqrt{2}-1/20$. 
	
	This example demonstrates what is our general expectation, namely that a generic extremal solution to the mixed correlator bootstrap will necessarily have ``particle production''. This just means that generically any operator that can appear in an OPE will do so. In particular, operators appearing in the 22 OPE must also appear in the 11 OPE and vice-versa. Intuitively this must be the case as to solve all the functional sum rules, in particular the mixed ones $\alpha^{1122}$, we necessarily need to turn on all possible OPE coefficients. There are in fact special corners in the space of solutions exist where this doesn't happen, GFF and minimal model correlators being prominent examples. It would be extremely interesting to find other ones.

	\widetext
 \clearpage

\end{document}